\documentclass[ams, prb, twocolumn, amssymb, floatfix, longbibliography]{revtex4-2}
\usepackage{amsmath}
\usepackage{tabularx}
\usepackage{bm}
\usepackage{euscript}
\usepackage{graphicx}
\usepackage[usenames,dvipsnames]{xcolor}
\usepackage{amsfonts}
\usepackage{exscale}
\usepackage{amsbsy}
\usepackage{subfigure}
\usepackage{textcomp}
\usepackage{comment}
\usepackage{slashed}
\usepackage{hyperref}
\usepackage{braket}

\hypersetup{
	colorlinks=true,
	linkcolor=black,     
	urlcolor=cyan,
	citecolor=ForestGreen
}

\newcommand{\nn}{\nonumber}

\newcommand{\beq}{\begin{equation}}
\newcommand{\eeq}{\end{equation}}
\newcommand{\be}{\begin{eqnarray}}
\newcommand{\ee}{\end{eqnarray}}

\begin{document}

\title{Possible continuous transition from fractional quantum Hall to stripe phase at $\nu_e=7/3$}
\author{Prashant Kumar$^{1}$ and R. N. Bhatt$^{2}$}
\affiliation{$^1$Department of Physics, Princeton University, Princeton NJ 08544, USA\\
		$^2$Department of Electrical and Computer Engineering, Princeton University, Princeton NJ 08544, USA
}
\date{\today}

\begin{abstract}
	We study the phase diagram of $\nu_e=7/3$ state in the $N=1$ Landau level in the presence of band mass anisotropy. Using density matrix renormalization group on an infinite cylinder geometry, we find a continuous transition from the topologically ordered Laughlin fractional quantum Hall state to a stripe phase with a period of approximately five and a half magnetic lengths. The transition is driven by the condensation of the magnetoroton mode which becomes gapless at the critical point. We interpret the transition within the composite-boson theory as the onset of stripe order in a superfluid background, resulting from the roton mode going soft.
\end{abstract}

\maketitle
\hypersetup{linkcolor=BrickRed}

\section{Introduction}


The two dimensional electron gas in a large perpendicular magnetic field exhibits a rich set of phenomena dominated by strong correlations between the electrons. If the applied magnetic field is strong, the electrons are essentially confined to Landau levels (LLs), with an integer number of filled LLs plus one fractionally filled LL at the top determined by the total filling fraction $\nu_e$. Since filled LLs give rise only to a quantized Hall conductance and are inert otherwise, this makes the interaction the sole term in the low energy Hamiltonian. 
Fractional quantum Hall (FQH) effect is a spectacular consequence of such interaction driven physics in the lowest few LLs.

An interesting question that arises is what phase transitions may occur between the FQH and other stable phases. In the presence of disorder, integer and fractional quantum Hall transitions are well known examples. Further, spontaneous nematic order may occur in higher LLs.\cite{Maciejko2013, You2014, Pan2014, Samkharadze2016, You2016, Regnault2017, Nguyen2018} In this paper, we find numerical evidence for a continuous transition from the Laughlin FQH to the stripe phase at $\nu_e=7/3$, when the Hamiltonian is translationally invariant but breaks the rotational symmetry.

The FQH states in $N=1$ LL have smaller gaps and are in general more susceptible to perturbations compared to the ones in the lowest LL.\cite{Wang2012,Yang2012,Rezayi2000,Faugno2021} In previous numerical studies, band anisotropy has been proposed to destabilize the FQH state in favor of a spontaneously broken translational symmetry state, in particular the stripe phase.\cite{Koulakov1996, Fogler1996, Fradkin1999,Yang2012,Papic2013,Zhu2017,He2021} However, the nature of such a transition is unclear. Motivated by this, we study the phase diagram of $\nu_e=7/3$ state as a function of anisotropy in the $N=1$ LL in the presence of Coulomb interactions.


We use the density matrix renormalization group on an infinite cylinder geometry (iDMRG) projected to the $N=1$ LL to study the effects of anisotropy.\cite{Mcculloch2008,Zaletel2015} By computing observables such as the correlation length, magnetoroton gap and the stripe order parameter, we show that the transition from the Laughlin FQH state to the stripe phase is continuous. The transition is driven by the condensation of the magnetoroton at a finite wavevector and it becomes gapless at criticality. We note that a continuous transition from a Wigner crystal to the stripe phase has been proposed to take place in the $N=2$ LL Ref. \onlinecite{He2021}.

In the Landau gauge, the onset of this phase is characterized by an $n$-fold increase in the periodicity of the Laughlin state and thus corresponds to the spontaneous breaking of $Z_n$ symmetry. It is found that $n$ increases with the circumference of the cylinder such that the period of the stripe phase remains unchanged when expressed in the units of magnetic length. 

The FQH and the stripe phase appear to be unrelated at first glance and a continuous transition between the two seems puzzling. We propose that the critical point may be understood from the perspective of the composite-boson (CB) theory.\cite{Zhang1989} The stripe phase can occur via the condensation of the roton mode of the CB superfluid and the transition can be second order.

This paper is organized as follows. In section \ref{sec:model}, we explain the model and the methods used to observe the FQH to stripe phase transition in iDMRG. We present evidence to support our claim that the transition is second order in section \ref{sec:exponents} and determine the critical exponents on the infinite cylinder geometry for the $Z_2,Z_3$ and $Z_4$ cases. 
We comment on the relation between the critical point in the quasi-1D geometry and $Z_n$ chiral clock models in section \ref{sec:chiral_clock_models}. The composite-boson interpretation of the transition is discussed in section \ref{sec:2D_limit}. We summarize our findings and give concluding remarks in section \ref{sec:Conclusions}. 

\section{Model and methods \label{sec:model}}
Let's consider electrons in two-dimensions interacting via a density-density interaction in the presence of a uniform perpendicular magnetic field $B$. In the limit when the interaction is weak compared to the cyclotron energy, it is sufficient to project to the highest energy LL that is partially filled. In this paper, we are interested in the $N=1$ LL. The electrons can be described by the following Hamiltonian:
\begin{align}
	H &= \frac{1}{2}\int \frac{d^2q}{(2\pi)^2}\ V(\bm q) f \left(\bm q\right)^2  :\bar\rho(\bm q)\bar\rho(-\bm q):\label{eq:Hamiltonian}\\
	\bar\rho(\bm q) &= \sum_{j} e^{-i\bm q.\bm R_j}
\end{align}
where $\ell^2 = 1/B$ is the magnetic length, $V(\bm q)$ is the interaction potential, $\bar\rho(\bm q)$ is the guiding center density operator and $f(\bm q)$ is the form factor of the $N=1$ LL.
Further, $\bm R_j$ are the guiding center coordinates of the $j^{th}$ particle. They are defined as $R^\alpha = r^\alpha + \ell^2\epsilon^{\alpha\beta}\Pi_\beta$ and satisfy the commutation relation $[R^\alpha,R^\beta]=-i\ell^2\epsilon^{\alpha\beta}$. Also, the kinetic momentum is $\Pi_\alpha \equiv p_\alpha-A_\alpha$ where $p_\alpha$ is the canonical momentum and $A_\alpha$ is the electromagnetic vector potential.
%
%

In our iDMRG simulations, we use an infinite cylinder geometry with its axis aligned along the $y$-direction and the periodic circumference direction along the $x$-axis. We consider the Landau gauge defined by the vector potential $\bm A = -By \hat x$. As a consequence of the translational invariance along $x$-direction, the single particle wavefunctions have the form: $\psi_k(\bm r) \propto e^{ikx} g(y+k\ell^2)$. Periodic boundary conditions along the circumference lead to $k =\kappa n$ where $\kappa\equiv 2\pi/L_x$, $L_x$ is the circumference of the cylinder and $n\in \mathbb{Z}$. Therefore, the LL orbitals in the Landau gauge form a 1D lattice with sites spaced by $2\pi/L_x$. The guiding center density and the Hamiltonian can be expressed in a second quantized form as follows:
\begin{align}
	\bar\rho(q_x, q_y) &= \delta_{q_x, \kappa k}\ e^{i\kappa k q_y/2} \sum_{n} e^{i\kappa n q_y} c^\dagger_n c_{n+k}\\
	H &= \frac{1}{2}\sum_{nmk} V_{mk}\ c_nc^\dagger_{n+k} c^\dagger_{n+m} c_{n+m+k}\label{eq:Hamiltonian_Vmk}\\
	V_{mk} &= \frac{1}{L_x} \int \frac{dq_y}{2\pi} \ V(\kappa k, q_y) f(\kappa k,q_y)^2 e^{i\kappa mq_y}
\end{align}
where $n,m,k\in \mathbb{Z}$ and only the anti-symmetric component of $V_{mk}$ (under $m\leftrightarrow k$) is important.

We introduce band anisotropy by considering an anisotropic mass tensor diagonal in the basis defined by the coordinate axes. To be precise, the kinetic energy has the following form before projection to the $N=1$ LL:
\begin{align}
	H_{K.E.} &= \frac{\Pi_x^2}{2m\alpha } + \frac{\alpha \Pi_y^2}{2m }
\end{align}
where $\alpha$ is a measure of the mass anisotropy. 
The  Hamiltonian when projected to the $N=1$ LL leads to an anisotropic form factor given by:
\begin{align}
	f(\bm q) &= \left(1-\frac{|q'|^2\ell^2}{2}\right)e^{-|q'|^2\ell^2/4}
\end{align}
where $q' \equiv q_x/\sqrt{\alpha} + i\sqrt{\alpha}q_y$.

In this paper, we assume that the electrons repel each other via isotropic Coulomb interactions, i.e.:
\begin{align}
	V(\bm r) &= \frac{e^{-r^2/2\xi^2}}{r}.\label{eq:Gaussian_coulomb}
\end{align}
where we have regulated the interaction at long distances using a Gaussian factor with a width equal to $\xi$. We take $\xi=6\ell$ which was found to not cause any significant effect in previous work.\cite{Krishna2019}

\subsection{Observing the stripe phase in quasi-1D geometry\label{sec:ground_state_method}}
In previous iDMRG simulations, the ground state at $\nu_e=7/3$, in the presence of an isotropic Coulomb potential, has been found to be the Laughlin FQH state with a longer correlation length compared to $\nu_e=1/3$.\cite{Zaletel2015} Therefore, unlike $\nu_e=1/3$,\cite{Wang2012} it could be more susceptible to a phase change. As we show later, the magnetoroton gap collapses upon increasing the band anisotropy and the system undergoes a phase transition to a stripe phase.

We observe the stripe phase in iDMRG as follows. In the infinite matrix-product state (iMPS) form, the Laughlin FQH ground state at $\nu_e=7/3$ is found in the root configuration $010$.\cite{Bernevig2008,Bergholtz2008} This implies that the ground state on the infinite cylinder can be constructed by repeating a unit cell MPS composed of three orbitals with the same quantum numbers as the electron configuration $010$. If the stripes in the stripe phase are parallel to the $x$-axis and its period contains $3n$ number of Landau orbitals, the translational symmetry of the Laughlin state is broken ``n''-fold. The transition to the ordered phase can then be observed by computing the ground state iMPS with a unit cell containing $3n$ orbitals at various values of anisotropy.

On the infinite cylinder, we denote the stripe order as $Z_n$. In general, ``n'' depends on the circumference. It is found that $n = \mathcal{O}(L_x/\ell)$, so that in the units of magnetic lengths, the period $a_0 = 3n \times 2\pi\ell^2/L_x$ remains roughly constant as we increase the cylinder circumference. This indicates that the stripe phase is not a finite-size artifact and is relevant to the 2D physics.

Our strategy for computing the ground state is as follows. We initialize the iMPS and environments with an electron configuration $0_n 1_n 0_n$ introducing an initial $Z_n$ stripe order. Here, the subscript denotes an $n$-fold repetition of the electron occupation number. Notice that this has the same quantum numbers as the configuration $(010)_n$ of the Laughlin FQH state. As we perform the iDMRG steps, strength of the stripe order parameter decays to zero (nonzero) if the system is in the FQH (stripe) state. We found good convergence with independent simulations at each anisotropy in the stripe phase. On the FQH side, we find that the best convergence and lowest energies are obtained if one starts in the FQH phase away from the transition and uses the converged ground state as an initial guess for a point closer to the transition. The ground state is thus adiabatically evolved as one moves towards the transition.

\subsection{Observables and critical scaling analysis}
Near the quantum phase transition, we measure observables and perform a critical scaling analysis. In particular, we measure three quantities. First, the correlation length ``$\xi$'' is calculated directly from the MPS form of the ground state using its transfer matrix.\cite{Schollwock2011} In general, all correlations decay faster compared to this length scale. We find that it is close to the one obtained by fitting the density-density correlation function to an exponentially decaying function.

Second, we calculate the order parameter ``$\rho_G$'' by measuring the density profile along the axial-direction of the cylinder and then taking a Fourier transform at the ordering wavevector $\bm G = (0,2\pi/a_0)$. Third, the magnetoroton gap ``$\omega_{\rm SMA}$'' at $\bm q = \bm G$ is measured using the single-mode approximation (SMA) of Girvin-Macdonald-Platzman.\cite{Girvin1986} While it is possible that SMA breaks down due to interactions between the gapless density fluctuations, it provides an energy scale for the critical fluctuations.\footnote{One such possible interaction can be between the magnetoroton at wavevector $\bm G$ and the state composed of three-rotons with a total momentum $\bm q = \bm q_1 + \bm q_2 + \bm q_3 = \bm G$, where $\bm q_1 = \bm q_2 = -\bm q_3 = \bm G$. Nevertheless, the SMA energy has a physical meaning as it is equal to the average energy of excitations weighted by the dynamical structure factor.\cite{Girvin1986} As long as the spectral weight of critical fluctuations diverges, it should capture the dynamical properties of critical fluctuations.} As such, it can be useful for studying dynamical scaling near the critical point.

We fit the three observables to power laws to obtain the critical exponents $\nu, \beta$ and $\nu z$ that correspond to the correlation length, order parameter strength and energy gap exponents respectively.\footnote{We fit the data using linear least squares error method. We have modified the algorithm by including weights proportional to the square of the observables. As long as the difference between the fitted curve and the data is small, this modification effectively corresponds to a linear least square error algorithm on a log-log scale.}

The iMPS obtained in iDMRG is an approximation of the true ground state. This is so because the iMPS truncates the infinite-dimensional Hilbert space of the cylinder to a finite one characterized by the bond-dimension $\chi$. One expects the observables to converge to the $\chi\rightarrow \infty$ case as $\chi$ is increased. However, we find that they do not show a numerically predictable behavior as a function of $\chi$. Therefore, we use the the largest $\chi$ we can access to perform the critical scaling analysis. We find that the finite bond-dimension affects the phase diagram in mainly two ways. First, the stripe phase is preferred since it has a lower entanglement entropy. Second, an iMPS with a finite bond-dimension can't have an infinite correlation length and a continuous transition becomes weakly first order. Fortunately, even though the observables show such dependence on $\chi$, the critical exponents are not altered significantly.

\section{Laughlin FQH to stripe phase transition at $\nu_e=7/3$ \label{sec:exponents}}
\begin{figure}
	\centering
	\includegraphics[width=0.35\textwidth]{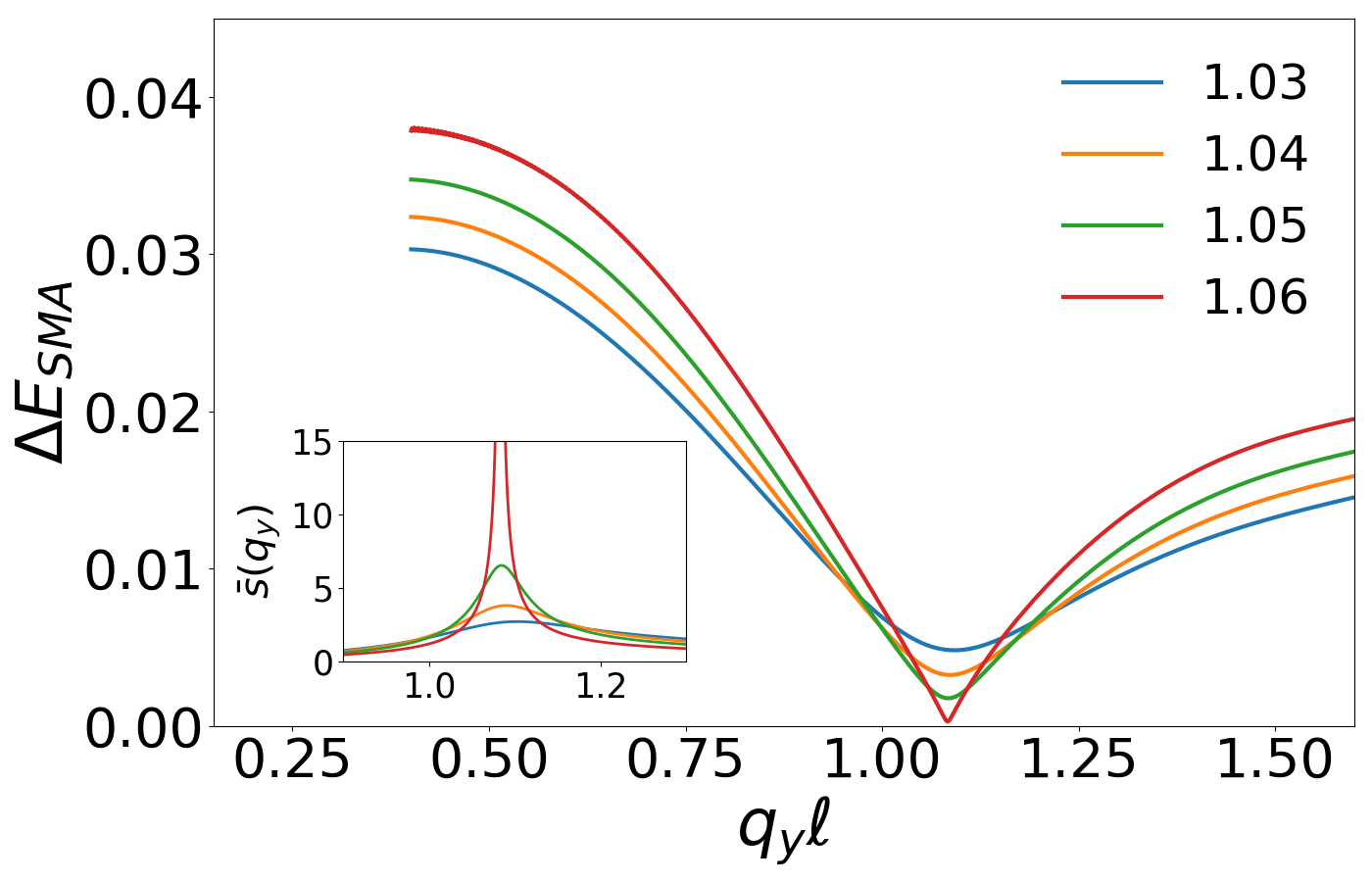}
	\caption{The magnetoroton spectrum computed using the single-mode approximation (SMA)\cite{Girvin1986} for various values of anisotropy ``$\alpha$'' (as indicated in the figure), $L_x = 6.5\ell$ and $q_y\ell = 0$. The SMA energy goes to zero at $q_y\ell = L_x/6\ell$ as the anisotropy is increased. This indicates the onset of stripe ordering with a period that contains $6$ Landau orbitals in the $y$-direction. The inset shows the divergence of the static structure factor $\bar s(q_y)$ as one approaches the transition.}
	\label{fig:roton_gap}
\end{figure}

In this section, we present our numerical results for the quantum phase transition from $\nu_e=7/3$ Laughlin FQH to the stripe phase as a band anisotropy is introduced in the system. We find that the transition is continuous, i.e., the inverse correlation length, energy gap and order parameter go to zero at the critical point. As evident in Fig. \ref{fig:roton_gap}, it is driven by the condensation of the magnetoroton which becomes gapless at criticality.

\begin{figure*}
	\centering
	\begin{minipage}{0.33\textwidth}
		\begin{center}
			\includegraphics[width=1.0\textwidth]{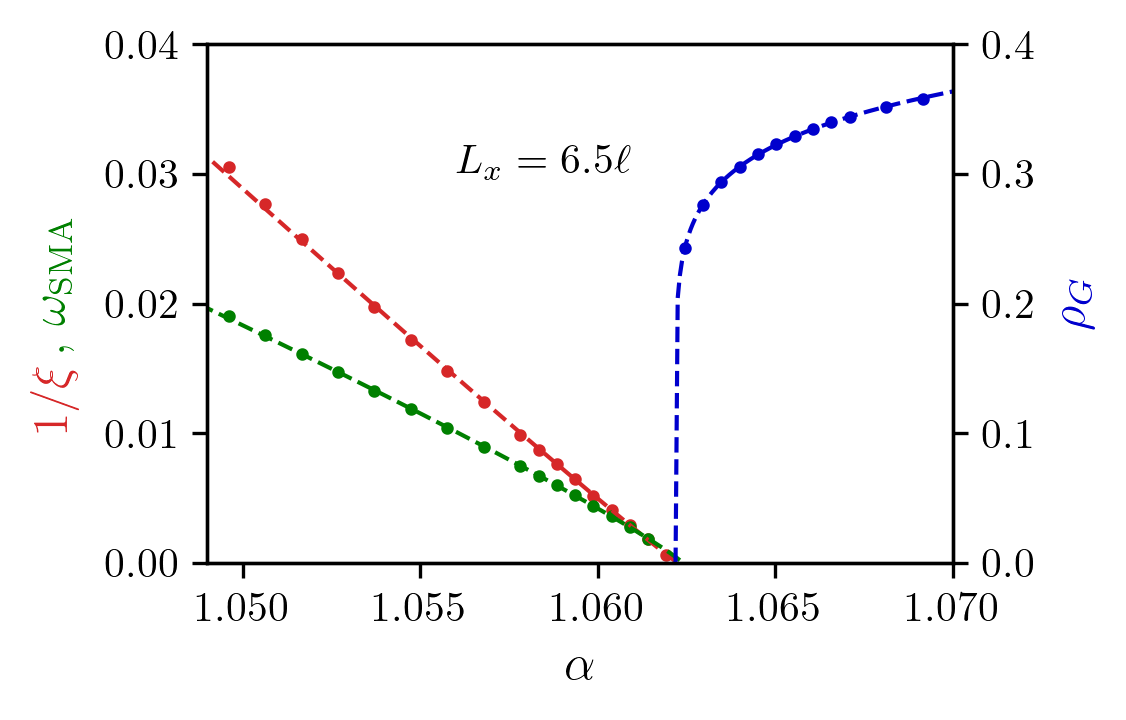}
			
			(a) $Z_2$
		\end{center}
	\end{minipage}%
	\begin{minipage}{0.33\textwidth}
		\begin{center}
			\includegraphics[width=1.0\textwidth]{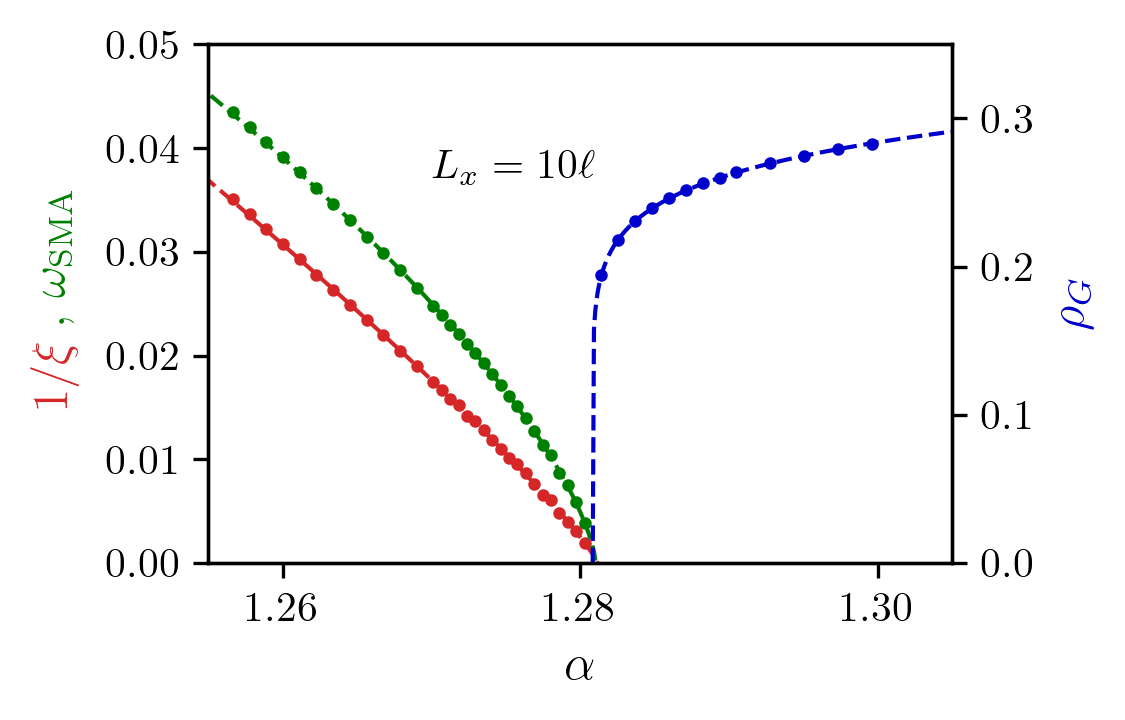}
			
			(b) $Z_3$
		\end{center}
	\end{minipage}%
	\begin{minipage}{0.33\textwidth}
		\begin{center}
			\includegraphics[width=1.0\textwidth]{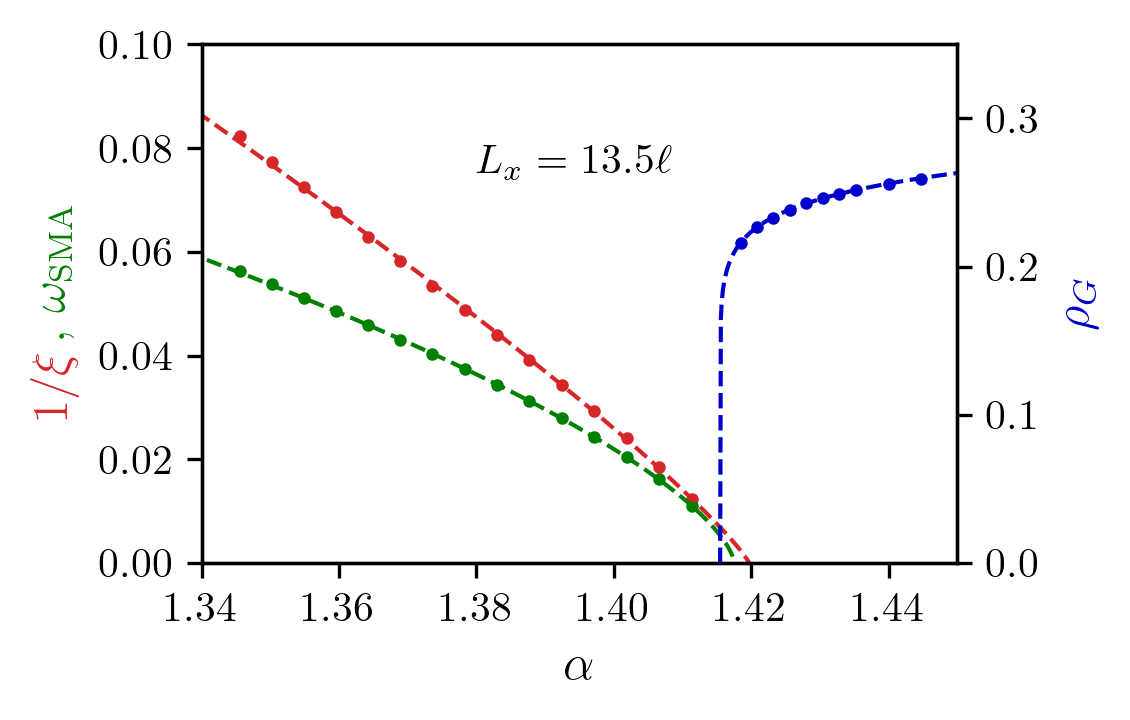}
			
			(c) $Z_4$
		\end{center}
	\end{minipage}
	
	\caption{Inverse correlation length ``$1/\xi$'' (red), SMA energy at the ordering wavevector ``$\omega_{\rm SMA}$'' scaled up by an appropriate factor (green) and order parameter ``$\rho_G$'' (blue) for Laughlin FQH to (a) $Z_2$, (b) $Z_3$ and (c) $Z_4$ stripe phase transition on the infinite cylinder. By fitting the three quantities to a power law, we obtain the exponents $(\nu, \nu z, \beta)$: (a) $(1.04, 0.90, 0.119)$, (b) $(0.87, 0.69, 0.109)$ and (c) $(0.86, 0.66, 0.08)$. The bond-dimensions $\chi$ are (a) $512$, (b) $4096$ and (c) $8192$. The fitted curves meet at a single point for $Z_2$ and $Z_3$ cases indicating a continuous transition. The appearance of weakly first order transition in $Z_4$ case may be due to stronger finite-size and bond-dimension effects.}
	\label{fig:exponents}
\end{figure*}

In Fig. \ref{fig:exponents}, we present the numerical results for the $Z_2, Z_3$ and $Z_4$ cases of the transition. The exponents $\nu$, $\nu z$ and $\beta$ are obtained by fitting the inverse correlation length, magnetoroton gap and order parameter to power laws respectively. Their values are reported in Table \ref{tab:table_of_exponents}. These should be interpreted as effective exponents rather than the asymptotic ones.\footnote{By ``effective exponents'', we mean the exponents obtained by fitting the data in the parameter regime accessible to our numerics. In general, the data may contain irrelevant corrections from scales such as the cylinder circumference $L_x$, interaction range $\xi$, bond-dimension $\chi$ and proximity to other critical points such as the achiral Potts model. On the other hand, the asymptotic exponents correspond to the limit in which these corrections vanish. To determine how close our effective exponents are to the asymptotic ones requires going closer to the critical point and bigger bond dimensions than we are currently able to.} Further, the energy gap exponent $\nu z$ depends on the validity of SMA. For the $Z_4$ case, we find that the bond-dimensions accessible to our numerical simulations are not sufficient to perform a reliable critical scaling analysis.\footnote{In addition, ground state at each anisotropy on the FQH side was obtained in an independent simulation as opposed to the adiabatic evolution method mentioned in section \ref{sec:ground_state_method}.} In particular, the correlation lengths are not large enough to avoid finite size effects that come from the finite cylinder circumference and interaction range. Nevertheless, we have indicated the exponents as obtained from the fitting procedure.

\begin{center}
	\begin{table}
		\begin{tabular}{ >{\centering\arraybackslash}p{2cm}|  >{\centering\arraybackslash}p{1cm} |  >{\centering\arraybackslash}p{1cm}| >{\centering\arraybackslash}p{1cm} }
			Symmetry & $\nu$ & $\nu z$ & $\beta$ \\\hline
			$Z_2$ & $1.04$ & $0.90$ & $0.119$ \\
			$Z_3$ & $0.87$ & $0.69$ & $0.109$ \\
			$Z_4$ & $0.86$ & $0.66$ & $0.08$ 
		\end{tabular}
		\caption{Critical exponents for the FQH to stripe phase transitions on an infinite cylinder.}
		\label{tab:table_of_exponents}
	\end{table}
\end{center}

As mentioned in section \ref{sec:ground_state_method}, generally, $n = \mathcal{O}(L_x/\ell)$. We find that the commensurate stripe phase is strongest when the relation $L_x \approx (3.5n-0.5)\ell$ is satisfied. Thus, we obtain the period of the stripe phase to be $a_0 \approx 5.4 \ell$.

\section{Quasi-1D critical point and $Z_n$ chiral clock models\label{sec:chiral_clock_models}}
Across the transition, the translation symmetry is spontaneously broken $n$-fold on the quasi-1D cylindrical geometry. We expect such a critical point to be described by the $Z_n$ chiral clock model.\cite{Ostlund1981,Huse1982,Fendley2012,Zhuang2015,Whitsitt2018} To see this explicitly, note that the different $Z_n$ stripe root configurations, for example $1_30_30_3$, $0_31_30_3$ and $0_30_31_3$ in the $Z_3$ case, can be mapped to an $n$-number of states of a clock model. In general, a domain wall configuration, that moves the clock clockwise when moving to the right, has a different energy compared to when the clock moves anticlockwise. Thus, the $Z_n$ clock model is chiral.\cite{Whitsitt2018}

Another way to see how the chirality arises is in terms of the discrete symmetries. 
The Hamiltonian of Eq. \eqref{eq:Hamiltonian_Vmk} has two discrete symmetries. First, there is an anti-unitary symmetry constructed by combining time-reversal with the mirror transformation that leaves the $y$-axis invariant. Additionally, we have an inversion symmetry which corresponds to a $180^0$ rotation of the quasi-1D cylinder. In the language of clock models,\cite{Zhuang2015} the first discrete symmetry corresponds to time-reversal while the second symmetry corresponds to the combination of charge conjugation and parity. Note that the charge conjugation and parity symmetries are broken individually in our problem which lead to the aforementioned chirality in the $Z_n$ clock model.\cite{Whitsitt2018} In this sense, the $n$-state Potts model which preserves all three discrete symmetries, i.e. charge conjugation, parity and time-reversal, is achiral.

Before comparing our results with the $Z_n$ clock models, we mention an important caveat. The Hamiltonian of Eq. \eqref{eq:Hamiltonian_Vmk} has an additional conservation law that restricts the domain wall configurations and may change the universality class. To see this, let's consider a thin torus geometry with $\ell \lesssim L_x \ll L_y$. The root configuration for the Laughlin FQH state is given by $(010)_{N_e}$, where $N_e$ is the number of electrons in the $N=1$ LL. In the Landau gauge, the total momentum along $x$-direction is conserved modulo $\kappa N_\phi$, where $N_\phi = 3N_e$. Thus, the momentum is:
\begin{align}
P_x/\kappa = \left(\sum_{k=0}^{N_\phi-1} k n_k\right)  \mod N_\phi
\end{align}
$n_k \equiv c^\dagger_kc_k$ is the occupation number operator of the $k^{th}$ orbital. Notice that the conservation of $P_x$ is equivalent to the conservation of center of mass along $y$-direction.

Assuming $N_e/n \in \mathbb{Z}$, there are $n$ degenerate ground states with the same total momentum $P_x$ in the stripe phase. These correspond to the stripe pattern $(0_n 1_n 0_n)_{N_e/n}$ and the ones generated when it is translated by $3m$ Landau orbitals, where $m \in \{1, 2, \cdots, n-1\}$. The domain walls between different ground states condense as one approaches the transition from the stripe phase side towards the FQH side. However, an arbitrary domain wall between two stripe patterns generally changes the position of the center of mass of electrons. Only a smaller set of configurations that respect the conservation law are allowed. At present, we are not aware how this restriction may change the universality class from the $\mathbb{Z}_n$ chiral clock model.

Fortuitously, the domain wall configurations for the $n=2$ case are unaffected. This is so because the two degenerate stripe patterns $001100$, $100001$ have exactly the same center of mass. As such, we expect the corresponding critical point to be in the universality class of the quantum Ising model in $1+1$ dimensions.
It predicts the following critical exponents: $\nu= \nu z=1,\ \beta=1/8$. These are within $5-10\%$ of the ones reported in Table \ref{tab:table_of_exponents} for the $Z_2$ case.

In $Z_3$ case, we first compare our results with the 3-state Potts model which corresponds to the achiral $Z_3$-symmetry breaking transition in one-dimension. It predicts $\nu = \nu z=5/6,\ \beta=1/9$. While $\nu$ and $\beta$ are close to what we report in Table \ref{tab:table_of_exponents}, $\nu z$ appears to be different. Since  we find the effective dynamical exponent $z$ to be less than one, the correlations travel faster compared to a light cone at long time scales and may be unphysical. This is similar to the $Z_2$ case and we believe that it is affected by the assumption of SMA. Nevertheless, the dynamical exponent is significantly different than the chiral 3-state clock model where $z>1$\cite{Samajdar2018, Chepiga2019, Nyckees2021}. A possible explanation is that our critical point is close to the achiral 3-state Potts point and thus contains crossover physics. Or, as explained earlier in this section, the restriction on the domain wall configurations, due to the conservation law, changes the universality class away from the $Z_3$ chiral clock model.

The achiral $Z_4$ transition corresponds to a family of universality classes described by the Ashkin-Teller model and has continuously varying exponents. Our exponent $\nu=0.86$ suggests that the chiral perturbation would be relevant.\cite{Schulz1983, Chepiga2021}  However, we note that the $Z_4$ case is also affected by the same caveats as the $Z_3$-symmetry breaking transition.

\section{Phase transition in the 2D limit \label{sec:2D_limit}}
The numerical results presented in this paper raise the interesting possibility of a continuous transition between a topologically ordered state and a conventional Landau ordered state, i.e., the stripe phase. At first sight, the two phases appear to be unrelated to one another. And the experience from Landau theory tells us that a direct transition between two unrelated conventionally ordered states is generally first order. As such, the nature of phase transitions that involve both the topological and Landau ordered states is an interesting problem. In this section, we argue that a continuous transition of the kind proposed in this paper can be understood within the framework of the composite-boson theory.\cite{Zhang1989}

Composite-bosons are emergent degrees of freedom obtained by attaching three flux-quanta to electrons. We assume that the energy scale associated with Coulomb interactions is weak compared to the cyclotron energy. Therefore, the physics of the $\nu_e=7/3$ state is essentially equivalent to a filling fraction $\nu_e=2+1/3 $, i.e, a $\nu_e=1/3$ state with the Hamiltonian given by Eq. \eqref{eq:Hamiltonian} and additional filled lowest LLs at $\nu_e=2$. As such, we attach flux quanta only to the electrons residing in the partially filled $N=1$ LL. The isotropic case can be described by the following lagrangian:
\begin{align}
\mathcal{L} &= i\phi^* D^a_t \phi + \mu \phi^*\phi - \frac{1}{2m} \phi^* (D^a_j)^2 \phi \nn\\
&\ \ \ \ \ \ \ \ + \frac{1}{12 \pi} (a-A)d(a-A) + \frac{1}{2\pi} AdA
\end{align}
where $\phi,\phi^*$ are the composite-boson fields and $D^a_\mu \equiv \partial_\mu -a_\mu$. $a_\mu$ and $A_\mu$ are the emergent and electromagnetic gauge fields respectively and $\mu$ corresponds to a chemical potential. The second to last term is the Chern-Simons term\footnote{$adb$ is a shorthand for $\epsilon^{\mu\nu\lambda} a_\mu\partial_\nu a_\lambda$} with its prefactor signifying the three attached flux-quanta. Additionally, the last term denotes the effect of the completely filled lowest LLs.

In the FQH side, the composite-boson ``$\phi$'' condenses and forms a superfluid. The Goldstone mode of the superfluid is eaten up by the gauge field and one obtains the $\nu_e=7/3$ FQH state with gapped excitations. Importantly, the superfluid has a collective roton mode\cite{feynman_1998} which produces the magnetoroton excitation above the FQH state.\cite{Girvin1986,Zhang1989} The transition to the stripe phase can take place if the roton condenses at a nonzero wavevector. The stripe phase would appear as coexisting superfluidity and the unidirectional CDW ordering. This is how a second order transition could be allowed within a Landau theory framework.

We briefly note that the transition can also be understood from the perspective of composite-fermions (CFs) formed by attaching two flux-quanta to electrons in the $N=1$ LL.\cite{Jain1989,Lopez91,Kalmeyer1992,Halperin1993,LeeJY2018} The $\nu_e=2+1/3$  filling fraction corresponds to a CF filling fraction $\nu_{\rm cf}= 1$. As such, the stripe phase may be obtained by the condensation of the exciton excitation of CFs at a finite wavevector corresponding to a missing CF in the lowest CF LL and an extra CF in the first CF LL.\cite{Jainbook}

\section{Conclusions \label{sec:Conclusions}}

In this paper, we have presented evidence for a direct continuous phase transition from the Laughlin fractional quantum Hall (FQH) state at $\nu_e=7/3$ to a stripe ordered phase. The transition is found to take place as one makes the system anisotropic by introducing a band anisotropy. 

We computed the correlation length, magnetoroton gap and the stripe order parameter on an infinite cylinder geometry and performed a critical scaling analysis. In a QH system on a cylinder, the translational symmetry is discrete and the stripe phase spontaneously breaks it ``n''-fold. In the $Z_2$ case, the exponents were found to be close to that of the quantum Ising model. However, for $Z_3$ and $Z_4$ cases, we find the exponents to be different compared to the chiral clock models, possibly due to the presence of the conservation of momentum along the circumference direction. Moreover, as one approaches the 2D limit by increasing the circumference, the effects of dislocations in the stripe order would become more important. These considerations suggests that the finite size scaling properties of the transition contain rich physics. We leave this as a problem to be addressed by future research.

Although the stripe and FQH states appear to be distinct, the theory of composite-bosons provides a framework to study the critical point between the two. We argued that stripe phase can be understood as a coexistence phase of composite-boson superfluid and a Bose-Einstein condensate of the roton. The theory of this 2D quantum critical point is an interesting problem. 

We note that the infinite cylinder geometry cannot distinguish between a smectic stripe phase and a stripe crystal.\cite{Fradkin1999} The latter breaks translational symmetry in both $x$- and $y$-directions and the order along the circumference direction is hard to detect. If the FQH state goes into the stripe crystal in the 2D limit, as the anisotropy is increased, we expect it to do it via a first order transition.

Experiments that study the effects of anisotropy are consistent with the emergence of a stripe order in the $N=1$ LL in the clean limit.\cite{Pan1999,Pan2001,Dean2008,Xia2010,Xia2011,Liu2013} However, it is unclear whether they see a nematic or a stripe phase since the translational order is difficult to observe. We believe that measuring the magnetoroton dispersion at finite wavevectors may provide a more direct evidence of the stripe phase.\cite{Kang2001,Rhone2011,Wurstbauer2015,Du2019,Liu2022}

\acknowledgments
We thank Srinivas Raghu, Steve Kivelson, Matteo Ippoliti and Hart Goldman for discussions. We thank the anonymous referee for providing insightful comments. The iDMRG numerical computations were carried out using the libraries developed by Roger Mong, Michael Zaletel and the TenPy collaboration. This work was supported by DOE BES Grant No. DE-SC0002140.

\bibliographystyle{utphys}
\bibliography{bigbib}

\providecommand{\href}[2]{#2}\begingroup\raggedright\begin{thebibliography}{10}

\bibitem{Maciejko2013}
J.~Maciejko, B.~Hsu, S.~A. Kivelson, Y.~Park, and S.~L. Sondhi, ``Field theory
  of the quantum hall nematic transition,''
  \href{http://dx.doi.org/10.1103/PhysRevB.88.125137}{{\em Phys. Rev. B}
  {\bfseries 88} (Sep, 2013) 125137}.
  \url{https://link.aps.org/doi/10.1103/PhysRevB.88.125137}.

\bibitem{You2014}
Y.~You, G.~Y. Cho, and E.~Fradkin, ``Theory of nematic fractional quantum hall
  states,'' \href{http://dx.doi.org/10.1103/PhysRevX.4.041050}{{\em Phys. Rev.
  X} {\bfseries 4} (Dec, 2014) 041050}.
  \url{https://link.aps.org/doi/10.1103/PhysRevX.4.041050}.

\bibitem{Pan2014}
W.~Pan, A.~Serafin, J.~S. Xia, L.~Yin, N.~S. Sullivan, K.~W. Baldwin, K.~W.
  West, L.~N. Pfeiffer, and D.~C. Tsui, ``Competing quantum hall phases in the
  second landau level in the low-density limit,''
  \href{http://dx.doi.org/10.1103/PhysRevB.89.241302}{{\em Phys. Rev. B}
  {\bfseries 89} (Jun, 2014) 241302}.
  \url{https://link.aps.org/doi/10.1103/PhysRevB.89.241302}.

\bibitem{Samkharadze2016}
N.~{Samkharadze}, K.~A. {Schreiber}, G.~C. {Gardner}, M.~J. {Manfra},
  E.~{Fradkin}, and G.~A. {Cs{\'a}thy}, ``{Observation of a transition from a
  topologically ordered to a spontaneously broken symmetry phase},''
  \href{http://dx.doi.org/10.1038/nphys3523}{{\em Nature Physics} {\bfseries
  12} no.~2, (Feb., 2016) 191--195},
  \href{http://arxiv.org/abs/1509.03658}{{\ttfamily arXiv:1509.03658
  [cond-mat.str-el]}}.

\bibitem{You2016}
Y.~You, G.~Y. Cho, and E.~Fradkin, ``Nematic quantum phase transition of
  composite fermi liquids in half-filled landau levels and their geometric
  response,'' \href{http://dx.doi.org/10.1103/PhysRevB.93.205401}{{\em Phys.
  Rev. B} {\bfseries 93} (May, 2016) 205401}.
  \url{https://link.aps.org/doi/10.1103/PhysRevB.93.205401}.

\bibitem{Regnault2017}
N.~Regnault, J.~Maciejko, S.~A. Kivelson, and S.~L. Sondhi, ``Evidence of a
  fractional quantum hall nematic phase in a microscopic model,''
  \href{http://dx.doi.org/10.1103/PhysRevB.96.035150}{{\em Phys. Rev. B}
  {\bfseries 96} (Jul, 2017) 035150}.
  \url{https://link.aps.org/doi/10.1103/PhysRevB.96.035150}.

\bibitem{Nguyen2018}
D.~X. Nguyen, A.~Gromov, and D.~T. Son, ``Fractional quantum hall systems near
  nematicity: Bimetric theory, composite fermions, and dirac brackets,''
  \href{http://dx.doi.org/10.1103/PhysRevB.97.195103}{{\em Phys. Rev. B}
  {\bfseries 97} (May, 2018) 195103}.
  \url{https://link.aps.org/doi/10.1103/PhysRevB.97.195103}.

\bibitem{Wang2012}
H.~Wang, R.~Narayanan, X.~Wan, and F.~Zhang, ``Fractional quantum hall states
  in two-dimensional electron systems with anisotropic interactions,''
  \href{http://dx.doi.org/10.1103/PhysRevB.86.035122}{{\em Phys. Rev. B}
  {\bfseries 86} (Jul, 2012) 035122}.
  \url{https://link.aps.org/doi/10.1103/PhysRevB.86.035122}.

\bibitem{Yang2012}
B.~Yang, Z.~Papi\ifmmode~\acute{c}\else \'{c}\fi{}, E.~H. Rezayi, R.~N. Bhatt,
  and F.~D.~M. Haldane, ``Band mass anisotropy and the intrinsic metric of
  fractional quantum hall systems,''
  \href{http://dx.doi.org/10.1103/PhysRevB.85.165318}{{\em Phys. Rev. B}
  {\bfseries 85} (Apr, 2012) 165318}.
  \url{https://link.aps.org/doi/10.1103/PhysRevB.85.165318}.

\bibitem{Rezayi2000}
E.~H. Rezayi and F.~D.~M. Haldane, ``{Incompressible Paired Hall State, Stripe
  Order, and the Composite Fermion Liquid Phase in Half-Filled Landau
  Levels},'' \href{http://dx.doi.org/10.1103/PhysRevLett.84.4685}{{\em Phys.
  Rev. Lett.} {\bfseries 84} (May, 2000) 4685}.

\bibitem{Faugno2021}
W.~N. Faugno, T.~Zhao, A.~C. Balram, T.~Jolicoeur, and J.~K. Jain,
  ``Unconventional ${\mathbb{z}}_{n}$ parton states at $\ensuremath{\nu}=7/3$:
  Role of finite width,''
  \href{http://dx.doi.org/10.1103/PhysRevB.103.085303}{{\em Phys. Rev. B}
  {\bfseries 103} (Feb, 2021) 085303}.
  \url{https://link.aps.org/doi/10.1103/PhysRevB.103.085303}.

\bibitem{Koulakov1996}
A.~A. Koulakov, M.~M. Fogler, and B.~I. Shklovskii, ``Charge density wave in
  two-dimensional electron liquid in weak magnetic field,''
  \href{http://dx.doi.org/10.1103/PhysRevLett.76.499}{{\em Phys. Rev. Lett.}
  {\bfseries 76} (Jan, 1996) 499--502}.
  \url{https://link.aps.org/doi/10.1103/PhysRevLett.76.499}.

\bibitem{Fogler1996}
M.~M. Fogler, A.~A. Koulakov, and B.~I. Shklovskii, ``Ground state of a
  two-dimensional electron liquid in a weak magnetic field,''
  \href{http://dx.doi.org/10.1103/PhysRevB.54.1853}{{\em Phys. Rev. B}
  {\bfseries 54} (Jul, 1996) 1853--1871}.
  \url{https://link.aps.org/doi/10.1103/PhysRevB.54.1853}.

\bibitem{Fradkin1999}
E.~Fradkin and S.~A. Kivelson, ``Liquid-crystal phases of quantum hall
  systems,'' \href{http://dx.doi.org/10.1103/PhysRevB.59.8065}{{\em Phys. Rev.
  B} {\bfseries 59} (Mar, 1999) 8065--8072}.
  \url{https://link.aps.org/doi/10.1103/PhysRevB.59.8065}.

\bibitem{Papic2013}
Z.~Papi\ifmmode~\acute{c}\else \'{c}\fi{}, ``Fractional quantum hall effect in
  a tilted magnetic field,''
  \href{http://dx.doi.org/10.1103/PhysRevB.87.245315}{{\em Phys. Rev. B}
  {\bfseries 87} (Jun, 2013) 245315}.
  \url{https://link.aps.org/doi/10.1103/PhysRevB.87.245315}.

\bibitem{Zhu2017}
Z.~Zhu, I.~Sodemann, D.~N. Sheng, and L.~Fu, ``Anisotropy-driven transition
  from the moore-read state to quantum hall stripes,''
  \href{http://dx.doi.org/10.1103/PhysRevB.95.201116}{{\em Phys. Rev. B}
  {\bfseries 95} (May, 2017) 201116}.
  \url{https://link.aps.org/doi/10.1103/PhysRevB.95.201116}.

\bibitem{He2021}
Y.~He, K.~Yang, M.~O. Goerbig, and R.~S.~K. Mong, ``Charge density waves and
  their transitions in anisotropic quantum hall systems,''
  \href{http://dx.doi.org/10.1038/s42005-021-00613-4}{{\em Communications
  Physics} {\bfseries 4} no.~1, (Jun, 2021) 116}.
  \url{https://doi.org/10.1038/s42005-021-00613-4}.

\bibitem{Mcculloch2008}
I.~P. McCulloch, ``Infinite size density matrix renormalization group,
  revisited,'' 2008.
\newblock \url{https://arxiv.org/abs/0804.2509}.

\bibitem{Zaletel2015}
M.~P. Zaletel, R.~S.~K. Mong, F.~Pollmann, and E.~H. Rezayi, ``Infinite density
  matrix renormalization group for multicomponent quantum hall systems,''
  \href{http://dx.doi.org/10.1103/PhysRevB.91.045115}{{\em Phys. Rev. B}
  {\bfseries 91} (Jan, 2015) 045115}.
  \url{https://link.aps.org/doi/10.1103/PhysRevB.91.045115}.

\bibitem{Zhang1989}
S.~C. Zhang, T.~H. Hansson, and S.~Kivelson, ``Effective-field-theory model for
  the fractional quantum hall effect,''
  \href{http://dx.doi.org/10.1103/PhysRevLett.62.82}{{\em Phys. Rev. Lett.}
  {\bfseries 62} (Jan, 1989) 82--85}.
  \url{https://link.aps.org/doi/10.1103/PhysRevLett.62.82}.

\bibitem{Krishna2019}
A.~Krishna, F.~Chen, M.~Ippoliti, and R.~N. Bhatt, ``Interaction-dependent
  anisotropy of fractional quantum hall states,''
  \href{http://dx.doi.org/10.1103/PhysRevB.100.085129}{{\em Phys. Rev. B}
  {\bfseries 100} (Aug, 2019) 085129}.
  \url{https://link.aps.org/doi/10.1103/PhysRevB.100.085129}.

\bibitem{Bernevig2008}
B.~A. Bernevig and F.~D.~M. Haldane, ``Model fractional quantum hall states and
  jack polynomials,''
  \href{http://dx.doi.org/10.1103/PhysRevLett.100.246802}{{\em Phys. Rev.
  Lett.} {\bfseries 100} (Jun, 2008) 246802}.
  \url{https://link.aps.org/doi/10.1103/PhysRevLett.100.246802}.

\bibitem{Bergholtz2008}
E.~J. Bergholtz and A.~Karlhede, ``Quantum hall system in tao-thouless limit,''
  \href{http://dx.doi.org/10.1103/PhysRevB.77.155308}{{\em Phys. Rev. B}
  {\bfseries 77} (Apr, 2008) 155308}.
  \url{https://link.aps.org/doi/10.1103/PhysRevB.77.155308}.

\bibitem{Schollwock2011}
U.~Schollwöck, ``The density-matrix renormalization group in the age of matrix
  product states,''
  \href{http://dx.doi.org/https://doi.org/10.1016/j.aop.2010.09.012}{{\em
  Annals of Physics} {\bfseries 326} no.~1, (2011) 96--192}.
  \url{https://www.sciencedirect.com/science/article/pii/S0003491610001752}.
  January 2011 Special Issue.

\bibitem{Girvin1986}
S.~M. Girvin, A.~H. MacDonald, and P.~M. Platzman, ``Magneto-roton theory of
  collective excitations in the fractional quantum hall effect,''
  \href{http://dx.doi.org/10.1103/PhysRevB.33.2481}{{\em Phys. Rev. B}
  {\bfseries 33} (Feb, 1986) 2481--2494}.
  \url{https://link.aps.org/doi/10.1103/PhysRevB.33.2481}.

\bibitem{Note1}
One such possible interaction can be between the magnetoroton at wavevector
  $\protect \bm {G}$ and the state composed of three-rotons with a total
  momentum $\protect \bm {q} = \protect \bm {q}_1 + \protect \bm {q}_2 +
  \protect \bm {q}_3 = \protect \bm {G}$, where $\protect \bm {q}_1 = \protect
  \bm {q}_2 = -\protect \bm {q}_3 = \protect \bm {G}$. Nevertheless, the SMA
  energy has a physical meaning as it is equal to the average energy of
  excitations weighted by the dynamical structure factor.\cite {Girvin1986} As
  long as the spectral weight of critical fluctuations diverges, it should
  capture the dynamical properties of critical fluctuations.

\bibitem{Note2}
We fit the data using linear least squares error method. We have modified the
  algorithm by including weights proportional to the square of the observables.
  As long as the difference between the fitted curve and the data is small,
  this modification effectively corresponds to a linear least square error
  algorithm on a log-log scale.

\bibitem{Note3}
By ``effective exponents'', we mean the exponents obtained by fitting the data
  in the parameter regime accessible to our numerics. In general, the data may
  contain irrelevant corrections from scales such as the cylinder circumference
  $L_x$, interaction range $\xi $, bond-dimension $\chi $ and proximity to
  other critical points such as the achiral Potts model. On the other hand, the
  asymptotic exponents correspond to the limit in which these corrections
  vanish. To determine how close our effective exponents are to the asymptotic
  ones requires going closer to the critical point and bigger bond dimensions
  than we are currently able to.

\bibitem{Note4}
In addition, ground state at each anisotropy on the FQH side was obtained in an
  independent simulation as opposed to the adiabatic evolution method mentioned
  in section \ref {sec:ground_state_method}.

\bibitem{Ostlund1981}
S.~Ostlund, ``Incommensurate and commensurate phases in asymmetric clock
  models,'' \href{http://dx.doi.org/10.1103/PhysRevB.24.398}{{\em Phys. Rev. B}
  {\bfseries 24} (Jul, 1981) 398--405}.
  \url{https://link.aps.org/doi/10.1103/PhysRevB.24.398}.

\bibitem{Huse1982}
D.~A. Huse and M.~E. Fisher, ``Domain walls and the melting of commensurate
  surface phases,'' \href{http://dx.doi.org/10.1103/PhysRevLett.49.793}{{\em
  Phys. Rev. Lett.} {\bfseries 49} (Sep, 1982) 793--796}.
  \url{https://link.aps.org/doi/10.1103/PhysRevLett.49.793}.

\bibitem{Fendley2012}
P.~Fendley, ``Parafermionic edge zero modes {inZn}-invariant spin chains,''
  \href{http://dx.doi.org/10.1088/1742-5468/2012/11/p11020}{{\em Journal of
  Statistical Mechanics: Theory and Experiment} {\bfseries 2012} no.~11, (Nov,
  2012) P11020}. \url{https://doi.org/10.1088/1742-5468/2012/11/p11020}.

\bibitem{Zhuang2015}
Y.~Zhuang, H.~J. Changlani, N.~M. Tubman, and T.~L. Hughes, ``Phase diagram of
  the ${Z}_{3}$ parafermionic chain with chiral interactions,''
  \href{http://dx.doi.org/10.1103/PhysRevB.92.035154}{{\em Phys. Rev. B}
  {\bfseries 92} (Jul, 2015) 035154}.
  \url{https://link.aps.org/doi/10.1103/PhysRevB.92.035154}.

\bibitem{Whitsitt2018}
S.~Whitsitt, R.~Samajdar, and S.~Sachdev, ``Quantum field theory for the chiral
  clock transition in one spatial dimension,''
  \href{http://dx.doi.org/10.1103/PhysRevB.98.205118}{{\em Phys. Rev. B}
  {\bfseries 98} (Nov, 2018) 205118}.
  \url{https://link.aps.org/doi/10.1103/PhysRevB.98.205118}.

\bibitem{Samajdar2018}
R.~Samajdar, S.~Choi, H.~Pichler, M.~D. Lukin, and S.~Sachdev, ``Numerical
  study of the chiral ${\mathbb{z}}_{3}$ quantum phase transition in one
  spatial dimension,'' \href{http://dx.doi.org/10.1103/PhysRevA.98.023614}{{\em
  Phys. Rev. A} {\bfseries 98} (Aug, 2018) 023614}.
  \url{https://link.aps.org/doi/10.1103/PhysRevA.98.023614}.

\bibitem{Chepiga2019}
N.~Chepiga and F.~Mila, ``Floating phase versus chiral transition in a 1d
  hard-boson model,''
  \href{http://dx.doi.org/10.1103/PhysRevLett.122.017205}{{\em Phys. Rev.
  Lett.} {\bfseries 122} (Jan, 2019) 017205}.
  \url{https://link.aps.org/doi/10.1103/PhysRevLett.122.017205}.

\bibitem{Nyckees2021}
S.~Nyckees, J.~Colbois, and F.~Mila, ``Identifying the huse-fisher universality
  class of the three-state chiral potts model,''
  \href{http://dx.doi.org/https://doi.org/10.1016/j.nuclphysb.2021.115365}{{\em
  Nuclear Physics B} {\bfseries 965} (2021) 115365}.
  \url{https://www.sciencedirect.com/science/article/pii/S0550321321000626}.

\bibitem{Schulz1983}
H.~J. Schulz, ``Phase transitions in monolayers adsorbed on uniaxial
  substrates,'' \href{http://dx.doi.org/10.1103/PhysRevB.28.2746}{{\em Phys.
  Rev. B} {\bfseries 28} (Sep, 1983) 2746--2749}.
  \url{https://link.aps.org/doi/10.1103/PhysRevB.28.2746}.

\bibitem{Chepiga2021}
N.~Chepiga and F.~Mila, ``Kibble-zurek exponent and chiral transition of the
  period-4 phase of rydberg chains,'' {\em Nature Communications} {\bfseries
  12} no.~1, (2021) 1--10. \url{https://doi.org/10.1038/s41467-020-20641-y}.

\bibitem{Note5}
$adb$ is a shorthand for $\epsilon ^{\mu \nu \lambda } a_\mu \partial _\nu
  a_\lambda $.

\bibitem{feynman_1998}
R.~P. Feynman, {\em Statistical mechanics: A set of lectures Chap. 11}.
\newblock Westview press, 1998.

\bibitem{Jain1989}
J.~K. Jain, ``{Composite-fermion approach for the fractional quantum Hall
  effect},'' \href{http://dx.doi.org/10.1103/PhysRevLett.63.199}{{\em Phys.
  Rev. Lett.} {\bfseries 63} (1989) 199}.

\bibitem{Lopez91}
A.~Lopez and E.~Fradkin, ``Fractional quantum hall effect and chern-simons
  gauge theories,'' \href{http://dx.doi.org/10.1103/PhysRevB.44.5246}{{\em
  Phys. Rev. B} {\bfseries 44} (Sep, 1991) 5246--5262}.
  \url{https://link.aps.org/doi/10.1103/PhysRevB.44.5246}.

\bibitem{Kalmeyer1992}
V.~Kalmeyer and S.-C. Zhang, ``{Metallic phase of the quantum Hall system at
  even-denominator filling fractions},''
  \href{http://dx.doi.org/10.1103/PhysRevB.46.9889}{{\em Phys. Rev. B}
  {\bfseries 46} (Oct, 1992) 9889--9892}.
  \url{http://link.aps.org/doi/10.1103/PhysRevB.46.9889}.

\bibitem{Halperin1993}
B.~I. Halperin, P.~A. Lee, and N.~Read, ``{Theory of the half-filled Landau
  level},'' \href{http://dx.doi.org/10.1103/PhysRevB.47.7312}{{\em Phys. Rev.
  B} {\bfseries 47} (Mar, 1993) 7312--7343}.
  \url{http://link.aps.org/doi/10.1103/PhysRevB.47.7312}.

\bibitem{LeeJY2018}
J.~Y. Lee, C.~Wang, M.~P. Zaletel, A.~Vishwanath, and Y.-C. He, ``Emergent
  multi-flavor ${\mathrm{qed}}_{3}$ at the plateau transition between
  fractional chern insulators: Applications to graphene heterostructures,''
  \href{http://dx.doi.org/10.1103/PhysRevX.8.031015}{{\em Phys. Rev. X}
  {\bfseries 8} (Jul, 2018) 031015}.
  \url{https://link.aps.org/doi/10.1103/PhysRevX.8.031015}.

\bibitem{Jainbook}
J.~K. Jain, {\em {Composite Fermions}}.
\newblock Cambridge University Press, 2007.

\bibitem{Pan1999}
W.~Pan, R.~R. Du, H.~L. Stormer, D.~C. Tsui, L.~N. Pfeiffer, K.~W. Baldwin, and
  K.~W. West, ``Strongly anisotropic electronic transport at landau level
  filling factor
  $\mathit{\ensuremath{\nu}}\phantom{\rule{0ex}{0ex}}=\phantom{\rule{0ex}{0ex}}9/2$
  and
  $\mathit{\ensuremath{\nu}}\phantom{\rule{0ex}{0ex}}=\phantom{\rule{0ex}{0ex}}5/2$
  under a tilted magnetic field,''
  \href{http://dx.doi.org/10.1103/PhysRevLett.83.820}{{\em Phys. Rev. Lett.}
  {\bfseries 83} (Jul, 1999) 820--823}.
  \url{https://link.aps.org/doi/10.1103/PhysRevLett.83.820}.

\bibitem{Pan2001}
W.~Pan, J.~Xia, E.~Adams, R.~Du, H.~Stormer, D.~Tsui, L.~Pfeiffer, K.~Baldwin,
  and K.~West, ``New results at half-fillings in the second and third landau
  level,'' \href{http://dx.doi.org/10.1016/s0921-4526(01)00284-8}{{\em Physica
  B: Condensed Matter} {\bfseries 298} no.~1-4, (2001) 113--120}.

\bibitem{Dean2008}
C.~R. Dean, B.~A. Piot, P.~Hayden, S.~Das~Sarma, G.~Gervais, L.~N. Pfeiffer,
  and K.~W. West, ``Contrasting behavior of the $\frac{5}{2}$ and $\frac{7}{3}$
  fractional quantum hall effect in a tilted field,''
  \href{http://dx.doi.org/10.1103/PhysRevLett.101.186806}{{\em Phys. Rev.
  Lett.} {\bfseries 101} (Oct, 2008) 186806}.
  \url{https://link.aps.org/doi/10.1103/PhysRevLett.101.186806}.

\bibitem{Xia2010}
J.~Xia, V.~Cvicek, J.~P. Eisenstein, L.~N. Pfeiffer, and K.~W. West,
  ``Tilt-induced anisotropic to isotropic phase transition at
  $\ensuremath{\nu}=5/2$,''
  \href{http://dx.doi.org/10.1103/PhysRevLett.105.176807}{{\em Phys. Rev.
  Lett.} {\bfseries 105} (Oct, 2010) 176807}.
  \url{https://link.aps.org/doi/10.1103/PhysRevLett.105.176807}.

\bibitem{Xia2011}
J.~Xia, J.~P. Eisenstein, L.~N. Pfeiffer, and K.~W. West, ``Evidence for a
  fractionally quantized hall state with anisotropic longitudinal transport,''
  \href{http://dx.doi.org/10.1038/nphys2118}{{\em Nature Physics} {\bfseries 7}
  no.~11, (Nov, 2011) 845--848}. \url{https://doi.org/10.1038/nphys2118}.

\bibitem{Liu2013}
Y.~Liu, S.~Hasdemir, M.~Shayegan, L.~N. Pfeiffer, K.~W. West, and K.~W.
  Baldwin, ``Evidence for a $\ensuremath{\nu}=5/2$ fractional quantum hall
  nematic state in parallel magnetic fields,''
  \href{http://dx.doi.org/10.1103/PhysRevB.88.035307}{{\em Phys. Rev. B}
  {\bfseries 88} (Jul, 2013) 035307}.
  \url{https://link.aps.org/doi/10.1103/PhysRevB.88.035307}.

\bibitem{Kang2001}
M.~Kang, A.~Pinczuk, B.~S. Dennis, L.~N. Pfeiffer, and K.~W. West,
  ``Observation of multiple magnetorotons in the fractional quantum hall
  effect,'' \href{http://dx.doi.org/10.1103/PhysRevLett.86.2637}{{\em Phys.
  Rev. Lett.} {\bfseries 86} (Mar, 2001) 2637--2640}.
  \url{https://link.aps.org/doi/10.1103/PhysRevLett.86.2637}.

\bibitem{Rhone2011}
T.~D. Rhone, D.~Majumder, B.~S. Dennis, C.~Hirjibehedin, I.~Dujovne, J.~G.
  Groshaus, Y.~Gallais, J.~K. Jain, S.~S. Mandal, A.~Pinczuk, L.~Pfeiffer, and
  K.~West, ``Higher-energy composite fermion levels in the fractional quantum
  hall effect,'' \href{http://dx.doi.org/10.1103/PhysRevLett.106.096803}{{\em
  Phys. Rev. Lett.} {\bfseries 106} (Mar, 2011) 096803}.
  \url{https://link.aps.org/doi/10.1103/PhysRevLett.106.096803}.

\bibitem{Wurstbauer2015}
U.~Wurstbauer, A.~L. Levy, A.~Pinczuk, K.~W. West, L.~N. Pfeiffer, M.~J.
  Manfra, G.~C. Gardner, and J.~D. Watson, ``Gapped excitations of
  unconventional fractional quantum hall effect states in the second landau
  level,'' \href{http://dx.doi.org/10.1103/PhysRevB.92.241407}{{\em Phys. Rev.
  B} {\bfseries 92} (Dec, 2015) 241407}.
  \url{https://link.aps.org/doi/10.1103/PhysRevB.92.241407}.

\bibitem{Du2019}
L.~Du, U.~Wurstbauer, K.~W. West, L.~N. Pfeiffer, S.~Fallahi, G.~C. Gardner,
  M.~J. Manfra, and A.~Pinczuk, ``Observation of new plasmons in the fractional
  quantum hall effect: Interplay of topological and nematic orders,'' {\em
  Science advances} {\bfseries 5} no.~3, (2019) eaav3407.
  \url{https://doi.org/10.1126/sciadv.aav3407}.

\bibitem{Liu2022}
Z.~Liu, U.~Wurstbauer, L.~Du, K.~W. West, L.~N. Pfeiffer, M.~J. Manfra, and
  A.~Pinczuk, ``Domain textures in the fractional quantum hall effect,''
  \href{http://dx.doi.org/10.1103/PhysRevLett.128.017401}{{\em Phys. Rev.
  Lett.} {\bfseries 128} (Jan, 2022) 017401}.
  \url{https://link.aps.org/doi/10.1103/PhysRevLett.128.017401}.

\end{thebibliography}\endgroup


\providecommand{\href}[2]{#2}\begingroup\raggedright\endgroup


\end{document}